\documentclass[sigconf]{acmart}
\AtBeginDocument{%
  }

\setcopyright{none}
\copyrightyear{}
\acmYear{2025}
\acmDOI{}
\setcopyright{none}
\renewcommand\footnotetextcopyrightpermission[1]{} 
\acmISBN{}

\usepackage{svg}
\usepackage[linesnumbered,ruled,vlined]{algorithm2e}
\usepackage{multirow}

\begin{document}

\title{Fast and Accurate Heuristics for Bus-Factor Estimation}

\author{Sebastiano A. Piccolo}
\email{sebastiano.piccolo@unical.it}
\orcid{0000-0002-6986-3344}
\affiliation{%
  \institution{University of Calabria \\ Department of Mathematics and Computer Science}
  \city{Rende (CS)}
  \country{Italy}
}

\renewcommand{\shortauthors}{Piccolo Sebastiano A.}

\begin{abstract}
The bus-factor is a critical risk indicator that quantifies how many key contributors a project can afford to lose before core knowledge or functionality is compromised. 
Despite its practical importance, accurately computing the bus-factor is NP-Hard under established formalizations, making scalable analysis infeasible for large software systems.

In this paper, we model software projects as bipartite graphs of developers and tasks and propose two novel approximation heuristics, Minimum Coverage and Maximum Coverage, based on iterative graph peeling, for two influential bus-factor formalizations. 
Our methods significantly outperform the widely adopted degree-based heuristic, which we show can yield severely inflated estimates.

We conduct a comprehensive empirical evaluation on over $1\,000$ synthetic power-law graphs and demonstrate that our heuristics provide tighter estimates while scaling to graphs with millions of nodes and edges in minutes. 
Our results reveal that the proposed heuristics are not only more accurate but also robust to structural variations in developer-task assignment graph.
We release our implementation as open-source software to support future research and practical adoption.
\end{abstract}

\maketitle

\section{Introduction}
Modern software systems are rarely built by isolated developers. 
Instead, they are shaped by large, collaborative ecosystems, often comprising hundreds or thousands of contributors working across geographies, time zones, and organizational boundaries \cite{yamashita2015pareto, majumder2019software}. 
While this collaborative model enables scalability and innovation, it also introduces serious risks: what happens if a critical contributor suddenly disappears from a project—leaves the company, switches teams, burns out, or simply stops responding?

This risk is commonly captured by the bus-factor, a colloquial yet increasingly formalized metric that quantifies the minimal number of people whose loss would severely stall project progress \cite{zazworka2010developers, avelino2016novel, rigby2016quantifying, cosentino2015assessing, piccolo2024evaluating}. 

The risk of knowledge loss or fragmentation due to people unavailability is particularly relevant in contexts such as open-source projects, where contributor commitment is voluntary and discontinuities are common \cite{majumder2019software, avelino2016novel} and in contexts where a small number of people are knowledgeable of most of the project, legacy components, and can be the gatekeepers to knowledge sharing \cite{piccolo2018design, lawrence1986organization, piccolo2022different, heath2000coordination, paloma2021beyond}.

Over the years, several formalizations of the bus-factor have been proposed. 
One line of work, including Avelino et al.\cite{avelino2016novel} and Zazworka et al.\cite{zazworka2010developers}, models projects as bipartite graphs between people and tasks (e.g., files, features, modules), and defines the bus-factor as the number of people whose removal leaves half the tasks uncovered. 
Another formulation, introduced by Piccolo et al.\cite{piccolo2024evaluating, piccolo2025busfactor}, focuses on the connectivity of tasks and defines the bus-factor in terms of how person removal affects the size of the largest connected component of tasks.

Despite these differences in definition, a shared obstacle emerges: computing the bus-factor is computationally hard under both formulations. 
In fact, even approximate solutions can be expensive on large graphs. 
As a result, practitioners often rely on simple heuristics, such as removing people in decreasing order of degree.

\begin{figure}
    \centering
    \includesvg[width=\linewidth]{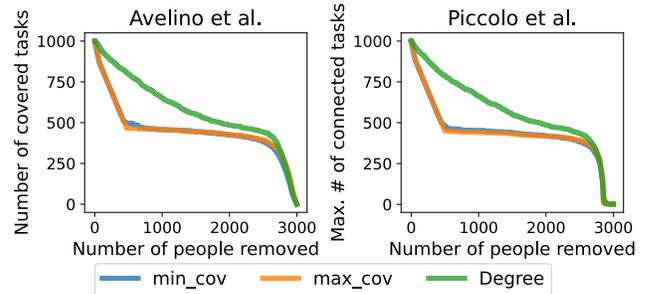}
    \caption{An example of bus-factor curves computed with both the definition from Avelino et al. and that from Piccolo et al. where the degree heuristics (green line) estimates a larger bus-factor than the one estimated by the two heuristics developed in this paper.}
    \Description{The image shows the inadequacy of the degree-based removal order heuristics.}
    \label{fig:example_degree_fails}
\end{figure}

Yet, as we show in Figure \ref{fig:example_degree_fails}, this approach can be dangerously misleading. 
In some cases, the degree heuristic grossly overestimates the number of people needed to reach critical task coverage thresholds, failing to prioritize people whose presence maintains key structural connectivity. The figure also shows the performance of our heuristics \texttt{Minimum Coverage} and \texttt{Maximum Coverage} which achieve significantly better estimates across both formulations.

These gaps in current practice raise several research questions:
\begin{itemize}
    \item Can we design heuristics that better approximate formal bus-factor definitions, while remaining scalable to large graphs?
    \item How do these heuristics behave under different structural properties of the bipartite graph of people and tasks?
    \item How well do these heuristics perform?
\end{itemize}

To address these questions, this paper makes the following contributions:
\begin{enumerate}
    \item We propose two new approximation heuristics: \texttt{Minimum Coverage} and \texttt{Maximum Coverage} which iteratively remove contributors with the minimum/maximum coverage to estimate the bus-factor more accurately.
    \item We introduce a greedy minimization strategy, which complements the aforementioned heuristics on the measure of Piccolo et al. by optimizing the structural disintegration of task connectivity.
    \item We provide a theoretical analysis of the time complexity of our algorithms and show how they scale to graphs with millions of nodes and edges.
    \item We conduct a comprehensive empirical evaluation on a large synthetic dataset of 1000 bipartite power-law graphs, showing that our heuristics outperform the state-of-the-art in both accuracy and robustness.
    \item We investigate how graph structure affects heuristic performance, providing new insight into when different heuristics succeed or fail.
\end{enumerate}

The remainder of the paper is organized as follows: section \ref{sec:related_work} reviews the measures of bus-factor which have been proposed so far, along with their estimation strategies.
Section \ref{sec:background} introduces the fundamentals of graph theory and formalizes the bus-factor problem along with the measures from Avelino et al. \cite{avelino2016novel}, Zazworka et al. \cite{zazworka2010developers}, and Piccolo et al. \cite{piccolo2024evaluating}.
Section \ref{sec:algorithms} describes the heuristics we have developed, provides their computational complexity and the relative implementation details.
Section \ref{sec:experiments} discuss our experimental evaluation where we assess the accuracy of our heuristics, their scalability, and their behavior under different graph structures.
Finally, section \ref{sec:conclusions} concludes the paper.

\section{Related work}
\label{sec:related_work}

The \textit{bus factor} (also known as \textit{truck factor}) quantifies the minimum number of key individuals whose sudden departure would harm a project. 
The bus-factor falls into the larger category of project issues known as \emph{community smells} \cite{almarimi2020learning,catolino2020refactoring,catolino2021understanding,paloma2021beyond}.
In software engineering research, this concept has been addressed primarily through two related problems: \emph{1)} identifying \textit{core developers}, those who contribute substantially to the codebase, and \emph{2)} estimating the project's bus factor given some model of contribution and task coverage.

A common method for identifying core developers is by computing a \textit{Degree of Authorship} (DoA), which assigns authorship scores to contributors for each file based on a variety of signals such as commit history, code reviews, or meetings~\cite{yamashita2015pareto, cosentino2015assessing, avelino2016novel, jabrayildaze2022bus, fritz2014degree}.
Typically, in order to compute its bus-factor, a codebase is represented as a bipartite graph of people and source code files and measures like the DoA are used to not only assess the contribution of each developer to each file, but also to filter the graph from non important contributions.

One of the earliest formalizations of the bus factor was introduced by Zazworka et al.~\cite{zazworka2010developers}, who proposed measuring project robustness as the number of people that can leave without jeopardizing task coverage beyond a specified threshold. 
Their framework defines three variants of the bus factor ($Z_{\min,t}$, $Z_{\max,t}$, and $Z_{\mathrm{avg},t}$), based on the minimum, maximum, and average number of tasks still covered after removing any $k$-sized subset of contributors. 
The general formulation is $Z_{\alpha,t} = \max \{n \in \mathbb{N} : \mathrm{cov}_\alpha(n) \ge tm\}$, where $m$ is the total number of tasks and $\mathrm{cov}_\alpha(n)$ measures coverage after removal.
Despite its generality, this approach has been criticized for being computationally infeasible for large projects~\cite{ricca2011difficulty, ferreira2017comparison} and for relying on fixed thresholds. Hannebauer and Gruhn~\cite{hannebauer2014algorithmic} showed that computing $Z_{\min,t}$ and $Z_{\max,t}$ is NP-hard, and although $Z_{\mathrm{avg},t}$ is polynomial-time computable, it is harder to interpret and less informative in practical scenarios.
More recently, Piccolo et al. \cite{piccolo2025busfactor} showed that $Z_{\max,t}$ is too optimistic and cannot capture variations to a project that increase/decrease its bus-factor (such as sparsifying the project graph by removing edges). 

To overcome the limitations of Zazworka’s exhaustive definitions, later work defined the bus-factor as the \emph{minimum} number of people whose removal causes the project to stall.
Cosentino et al.~\cite{cosentino2015assessing} introduced a commit-based tool that identifies primary developers (who meet a high knowledge threshold $X$, e.g. $1/N$ of total authorship) and secondary developers (above a lower threshold $Y$, often $Y=0.5X$). The bus factor is then the size of the union of these two sets of developers \cite{lisan2024guiding}.
This two-threshold scheme can highlight the few experts covering a component, but it requires choosing $X$ and $Y$ a priori and can be sensitive to those parameters.
Avelino et al.~\cite{avelino2016novel} proposed a simpler DOA-based heuristic: compute a DoA score for each developer on each file, then iteratively remove the current top author (the one covering the most files) and increment the bus-factor count, stopping when fewer than 50\% of files remain covered by any author.
In other words, one keeps removing people until more than half the code would become abandoned~\cite{avelino2016novel}. 
Rigby et al.~\cite{rigby2016quantifying}, coming from a risk management perspective, considered both the likelihood and the impact of a loss of developers and proposed a heuristic based on a Monte Carlo simulation of personnel loss to estimate the bus-factor of a project.
Ferreira et al.\cite{ferreira2017comparison}, in a comparison between bus-factor estimation heuristics, found that Avelino’s removal algorithm and Cosentino’s approach were the most accurate predictors of bus factor (and key developers), with Avelino’s method performing slightly better on average.
Recently, Jabrayilzade et al. \cite{jabrayildaze2022bus} proposed a multimodal bus-factor estimator that extends Avelino’s DoA metric by decaying contributions over time and including review/meeting metadata.
Their algorithm slightly outperformed Avelino’s original DOA-only tool at predicting both the bus factor and the key engineers identified by survey respondents \cite{jabrayildaze2022bus}.

Machine learning approaches have also emerged to predict bus-factor risk. Almarimi et al.~\cite{almarimi2020learning, almarimi2023improving} developed \textsc{csDetector}, which flags the presence of a "truck-factor smell"—a form of community smell—using decision tree classifiers. 
While effective at identifying risky projects, \textsc{csDetector} does not estimate a numeric bus factor, instead requiring one as input for training. 
Its binary nature makes it more suitable as a complementary diagnostic tool.

To overcome the limitations of threshold-dependence and non-normalized metrics, Piccolo et al.~\cite{piccolo2018design, piccolo2024evaluating, piccolo2025busfactor} recently proposed a graph-theoretic, threshold-free formulation of the bus factor. 
They represent a project as a bipartite graph between people and tasks and define the bus factor as the normalized area under the \textit{robustness curve}, which measures how the largest connected component of tasks degrades as contributors are removed in adversarial order. 
This normalized metric facilitates comparison across projects of varying size and complexity. 
Although computing the exact metric is NP-hard\cite{piccolo2025busfactor}, the authors provide a linear-time approximation algorithm based on union-find structures and propose hill-climbing and simulated annealing heuristics for optimizing team-task assignments. 
They demonstrated that their method can improve a real project's bus factor by over 40\% through strategic reallocation.

In summary, bus-factor estimation has evolved from fixed threshold, coverage-based metrics to more nuanced, scalable, and generalizable approaches, which include also machine learning approaches and normalized application-agnostic measures. 
This paper joins this body of knowledge by contributing with fast and scalable heuristics to estimate the bus-factor for the Avelino et al. \cite{avelino2016novel} and Piccolo et al. \cite{piccolo2024evaluating, piccolo2025busfactor} measures.

\section{Background}
\label{sec:background}
In this paper, we use bipartite graphs as tools to model software projects and estimate their bus-factor.
In section \ref{sub:graph_theory}, we introduce the graph theoretical tools and the notation that we will use throughout the paper; in section \ref{sub:bus_factor}, we introduce the bus-factor problem, its formalizations, and the current estimation technique.

\subsection{Graph theory}
\label{sub:graph_theory}

\subsubsection{Preliminaries}
A graph is a pair $\mathcal{G} = (V, E)$ where $V$ is the set of vertices and $E \subseteq V \times V$ is the set of edges. 
A graph is represented through its adjacency matrix $A$; with $A_{ij} = 1$ if vertices $i$ and $j$ are connected and $A_{ij} = 0$ otherwise. 
The degree $k_i$ of a vertex $i$ is the number of vertices connected to it; that is, $k_i = \sum_j A_{ij}$.
A \emph{subgraph} of a graph $G$ is the graph induced by a subset $S \subseteq V$ of vertices of $G$ and it is denoted by $G[S]$.
A connected component of a graph is a connected subgraph which is not part of any other connected subgraph.

\subsubsection{Bipartite graphs and modeling of software projects}
If the set of vertices $V$ can be divided in two sets $V_1$ and $V_2$ with $V_1 \cap V_2 = \emptyset$ and $V_1 \cup V_2 = V$, such that $\forall (i,j) \in E \; i \in V_1 \wedge j \in V_2$, then the graph is called \emph{bipartite}. 
A bipartite graph is the appropriate tool to represent the connections between a set of people and a set of artifacts such as source code files, documents, modules, tasks, and so on.
Without loss of generality, in this paper we refer to bipartite graphs of people and tasks, which we indicate as $\mathcal{G} = (P, T, E)$, where $P$ is the set of people, $T$ is the set of tasks, and $E$ is the set of edges between people ans tasks.
We say that a person $i$ \emph{covers} a task $j$ if $(i,j) \in E$ and we say that a task $j$ is covered if there is at least a person connected to it; that is if $k_j > 0$.
We refer to the tasks in a connected component as \emph{connected tasks} and we refer to the largest number of tasks in a connected component as the \emph{maximum number of connected tasks}, and denote it with $\tau(G)$.
Finally, $k_P = \max_{i \in P} \sum_i A_{ij}$ and $k_T = \max_{j \in T} \sum_j A_{ij}$ indicate, respectively, the largest degree among the people and the largest degree among the tasks.

\subsubsection{Power-law graphs and heroes in software projects}
Many real-world graphs exhibit right skewed degree distributions where a small fraction of nodes has high degree while the vast majority of nodes has small degree.
Software and engineering projects are no exception, and highly uneven workload and knowledge distributions, among the people involved, have been empirically determined.
In software projects, people who take on a large number of tasks compared to the remaining people are often referred to as \emph{heroes}.

A convenient way to generate graphs with highly skewed degree distributions is to sample node degrees from a power-law distribution, connecting nodes randomly according to the sampled degree distribution.
A power-law distribution is defined by the following probability density function:
\begin{equation}
    \label{eq:power_law}
    f(x, \lambda, k_{\min}, k_{\max}) = \frac{\lambda}{k_{\max}}\left( \frac{x - k_{\min}}{k_{\max}} \right)^{\lambda - 1}
\end{equation}
where $0 < \lambda < 1$ is a parameter that regulates the skewness of the distribution: the higher $\lambda$ the less skewed the distribution (if $\lambda = 1$ we obtain the uniform distribution); $k_{\min}$ and $k_{\max}$ regulate the support of $f$, which is equal to $[k_{\min}, k_{\min} + k_{\max}]$, and therefore the maximum and minimum degree in a power-law graph.

Algorithm \ref{alg:bipartite_power_law_graph} provides a procedure to create a bipartite power-law graph.
It works as follows: first, it samples two two power-law degree distributions, with no isolated node (no node has degree 0).
Then, it ensures that the two degree distributions have the same sum and generates a graph by connecting random nodes according to the designated degrees, and without creating double edges between the same couple of nodes.
Finally, if the graph is not connected, performs a random rewiring procedure which does not alter the degree distributions.

\begin{algorithm}
\SetAlgoLined
\KwIn{$|P|$, $|T|$, $\lambda_P$, $\lambda_T$, $k_P$, $k_T$}
\KwOut{A connected power-law bipartite graph $G$}
\BlankLine

$\mathrm{degrees}_P \sim \mathrm{PowerLaw}(\lambda_P, 1, k_P-1, |P|)$\;
$\mathrm{degrees}_T \sim \mathrm{PowerLaw}(\lambda_T, 1, k_T-1, |T|)$\;

\While{$\sum \mathrm{degrees}_P \neq \sum \mathrm{degrees}_T$}{
    Select a random node with degree > 1 from the larger distribution and reduce its degree by one\;
}

$G \gets$ ConfigurationModel($\mathrm{degrees}_P$, $\mathrm{degrees}_T$)\;

\While{$G$ \textbf{is not} connected}{
    Choose two connected components $C_1, C_2$ from $G$\;
    Choose two edges: $(p_1, t_1)$ from $C_1$ and $(p_2, t_2)$ from $C_2$\;
    Rewire the edges as follows: $(p_1, t_2),(p_2, t_1)$;
}

\Return $G$\;

\caption{Bipartite Power-Law Graph Generator}
\label{alg:bipartite_power_law_graph}
\end{algorithm}

\subsection{Bus-Factor}
\label{sub:bus_factor}

The bus-factor is a measure of project vulnerability against personnel loss and is defined, informally, as the number of people that have to leave a project -- as if they were hit by a bus -- until the project stalls.
Defining when a project stalls is far from easy and different bus-factor measures define the stalling condition in different ways.
Here, we introduce the bus-factor problem and the existing approaches to estimate it.

\begin{definition}[The bus-factor problem]
Let $G = (P, T, E)$ be a graph representing a project of $|P| = n$ people and $|T| = m$ tasks.
Let $S \subseteq P$ be the smallest set of people whose loss would cause the project to stall.
The bus-factor of $G$, denoted with $B(G)$, is given by the cardinality of $S$: $B(G) = |S|$.
The bus-factor can also be expressed in relative terms, if divided by the theoretical maximum.
\end{definition}

What remains to be defined from the previous definition is the project stalling condition.
This is far from trivial and possibly domain dependent. 
To date, measures of bus-factor define the stalling condition either in terms of task coverage or in terms of maximum number of tasks connected in a single component.

\subsubsection{The Avelino et al. measure of bus-factor}
The most common measure of bus-factor comes from Avelino et al. who assumed that a project stalls if the number of covered tasks is less than $t \times m$, where $0 < t < 1$ is an input parameter.
Let $\mathrm{cov}(G) = |\{j: k_j > 0, \forall j \in T\}|$ be the number of covered tasks in $G$, bus-factor according to Avelino et al. is as follows:

\begin{equation}
    A(G, t) = \min_{S \subseteq P} |S|, \quad \mathrm{cov(G[P \setminus S])} < t \times m
    \label{eq:avelino}
\end{equation}

In other words, $A(G)$ is the size of smallest subset of people $S$ that, if removed from $G$, causes more than $t \times m$ tasks to become isolated.
Avelino et al. suggest to set $t = 0.5$.

A related measure of bus-factor, which has been used in the past before the introduction of (\ref{eq:avelino}), is due to Zazworka et al. and is defined as follows:

\begin{equation}
    Z(G, t) = \arg \max_{k \in \mathbb{N^+}} \min_{S \subseteq P, |S| = k} \mathrm{cov(G[P \setminus S])} \ge t \times m
    \label{eq:zazworka}
\end{equation}

that is, the largest subset $S$ of people such that, if removed from $G$, the minimum number of covered tasks is greater or equal to $t \times m$.

\begin{lemma}
    $Z(G, t) = A(G, t) - 1$.
\end{lemma}

\begin{proof}
    Since $A(G, t)$ is the size of the smallest set $S$ such that $\mathrm{cov(G[P \setminus S])} < t \times m$, for any set $X \subseteq P$ of size $|S| - 1$ it holds $\mathrm{cov(G[P \setminus X])} \ge t \times m$, and $|S| - 1$ is the largest possible size for such a set $X$. As such, $Z(G, t) = A(G, t) - 1$.
\end{proof}
This lemma implies that a solution to (\ref{eq:avelino}) is also a solution to (\ref{eq:zazworka}).
However, solving (\ref{eq:avelino}) is known to be NP-Hard; therefore, in practice, we use an approximation algorithm to estimate it.
The heuristic suggested by Avelino et al. is to remove people in decreasing order of their degree, until the number of covered tasks is lower than 50\%.
The number of the people removed in this way is the estimated bus-factor.

\subsubsection{The Piccolo et al. measure of bus-factor}
Recently, Piccolo et al. proposed a novel measure of bus-factor which is normalized and does not depend on an input threshold \cite{piccolo2024evaluating, piccolo2025busfactor}.
They argue that the use of threshold is somewhat arbitrary and that the task coverage does not capture the degree of knowledge fragmentation in a project as people leave.
For instance, a project consisting of $n$ disconnected dyads (that is, and edge connecting one person with one task) has all the tasks covered, but Piccolo et al. \cite{ piccolo2025busfactor} argue that it has a high degree of fragmentation and that this should be accounted for by a bus-factor measure.

Consequently, they substituted the coverage with the maximum number of connected tasks; that is, the largest number of tasks contained in a connected component.
In order not to rely on a threshold, they proposed to compute the area under the decay curve of the maximum number of connected tasks as people are removed, normalizing it by the theoretical maximum (obtained for a corresponding fully connected bipartite graph).
Therefore, they define the bus-factor as the minimum area under the curve over all possible people removal orders \cite{piccolo2024evaluating, piccolo2025busfactor}.

Formally, let $\pi: P \rightarrow P$ be a permutation of the people in $P$, $\Pi(P)$ be the space of all permutations of $P$, and $G_i^{\pi} = G[P \setminus \{ \pi_1, \pi_2, \dots, \pi_i\}]$ be the residual graph obtained by removing the all the people up to the i-th person in the permutation $\pi$.
Recall that $\tau(G)$ is the maximum number of tasks connected into a component of $G$.
The bus-factor according to Piccolo et al. \cite{piccolo2025busfactor} is defined as follows:

\begin{equation}
    \mathcal{P}(G) = \min_{\pi \in \Pi(G)} \frac{1}{2n - 1}\sum_{i = 1}^n \tau(G_{i-1}^{\pi}) + \tau(G_{i}^{\pi})
    \label{eq:piccolo}
\end{equation}

where $G_0^{\pi} = G$.
Note that the term $\sum_{i = 1}^n \tau(G_{i-1}^{\pi}) + \tau(G_{i}^{\pi})$ in (\ref{eq:piccolo}) is just the standard trapezoid rule for computing the area under the curve and the denominator ${2n - 1}$ is the maximum theoretical value, obtained for a fully connected graph with $n$ people.
As such, $\mathcal{P}(G)$ takes values in $[0,1]$.

Piccolo et al. \cite{piccolo2025busfactor} proved that solving (\ref{eq:piccolo}) is NP-Hard and therefore they suggest to approximate it by ordering the people in $\pi$ in decreasing order of degree.

\section{New heuristics for bus-factor estimation}
\label{sec:algorithms}
In this section, we present the heuristics we developed in order to achieve a superior bus-factor estimation.
Specifically, we develop two graph peeling heuristics.
The strategy described in section \ref{sub:min_cov}, named \texttt{Minimum Coverage}, iteratively selects the person which covers the smallest number of tasks.

The strategy described in section \ref{sub:max_cov}, named \texttt{Maximum Coverage}, iteratively selects the person which covers the largest number of uncovered tasks.
These strategies work very well for both the Avelino et al. measure, $A(G, t)$, and the Piccolo et al. measure, $\mathcal{P}(G)$.

Finally, in section \ref{sub:reverse_greedy} we introduce a heuristic which iteratively selects the person which grows $\tau(G)$ by the smallest number of tasks.
This is specifically design to address $\mathcal{P}(G)$ and is used in combination with \texttt{Minimum Coverage} and \texttt{Maximum Coverage} in order to improve the estimation of $\mathcal{P}(G)$.

Our heuristics take a graph $G = (P,T,E)$ in input and output a people removal order $\pi$ which can be used to estimate both $A(G, t)$ and $\mathcal{P}(G)$.

\subsection{Minimum Coverage Iterative Peeling}
\label{sub:min_cov}
The Minimum Coverage Iterative Peeling heuristic is a greedy algorithm for producing an ordering of nodes in a bipartite graph that prioritizes minimal task coverage.
The goal is to construct an ordering $\pi$ of people in $P$ such that people covering fewer \emph{currently} uncovered tasks appear earlier in $\pi$.

The heuristic is motivated by the idea of preserving the most \emph{valuable} people, those who cover the most yet-uncovered tasks, for as long as possible in the ordering. 
By iteratively removing the person with the smallest marginal contribution to the set of uncovered tasks, we aim to delay the coverage of many tasks. 

In this way, reversing the resulting order $\pi$ provides a sequence in which tasks become uncovered as rapidly as possible upon the removal of people, making it a powerful heuristic for bus-factor estimation.

\subsubsection{Algorithm description}
We start with an empty ordering $\pi$.
At each step, for each person $p \in P \setminus \pi$, we define their current coverage as the number of tasks they cover that are not yet covered by the people in $\pi$. 
The person with the smallest current coverage is appended to $\pi$. 
This process continues until all people have been added to the ordering $\pi$.

Equivalently, one can view this process as an incremental construction: starting with a graph that includes only the task nodes, people are added one at a time in order of least contribution to task coverage. 
This greedy strategy ensures that tasks are covered as slowly as possible, at each step, preserving highly covering individuals for later stages.

\subsubsection{Implementation details}
A naive implementation of \texttt{Minimum Coverage} would require to recompute the effective task coverage for each person connected to at least one task covered by the last person added to $\pi$.
Updating the effective coverage of the affected people requires $O(|E|)$ operations.
Finding the person with the minimum coverage requires $O(|P|)$
Since we have $|P|$ people, the total computational complexity of the naive implementation would be $O(|E| \times |P|) = O(|P|^2 \times |T|)$, which would not scale to large graphs.

\begin{algorithm}
\KwIn{a bipartite graph $G = (P, T, E)$}
\KwOut{Ordering $\pi$ of people}

$\texttt{inserted[$p$]} \gets 0, \; \forall p \in P $\;

$\mathcal{Q}$ $\gets$ MinHeap()\;
\ForEach{$p \in P$}{
    $\mathcal{Q}$.push($(G.\text{degree}(p), p)$)\;
}

$\pi \gets []$\;

\While{$\mathcal{Q}$ is not empty}{
    $(\texttt{priority}, p) \gets \mathcal{Q}.\text{pop()}$\;
    
    \If{\texttt{inserted[$p$]}}{
        \textbf{continue}\;
    }

    $\pi$.append($p$)\;
    \texttt{inserted[$p$]} $\gets$ 1\;

    $\texttt{covered\_tasks} \gets G.\text{neighbors}(p)$\;
    $\texttt{affected} \gets \bigcup_{t \in \texttt{covered\_tasks}} G.\text{neighbors}(t) \setminus \{p\}$\;

    $G.\text{remove\_nodes\_from}(\texttt{covered\_tasks})$\;

    \ForEach{$q \in \texttt{affected}$}{
        $\mathcal{Q}$.push($(G.\text{degree}(q), q)$)\;
    }
}
\Return{$\pi$.\textnormal{reverse($\,$)}}\;

\caption{Minimum Coverage}
\label{alg:min_cov}
\end{algorithm}

Instead, we adopt a priority queue, where people are prioritized by effective coverage (Algorithm \ref{alg:min_cov}, lines 2---4).
We iteratively pop the person $p$ with the minimum coverage and, if it is not a person we have already processed, we add it to $\pi$ (lines 7---10).
We mark the person $p$ as \emph{inserted}, obtain the set of covered tasks, and the set of affected people, i.e. people who are connected to at least one newly covered task (lines 11---13).
We remove the covered tasks from the graph $G$ and update the priority, i.e. the effective coverage, for the affected people (lines 14---16).
When the algorithm terminates, we return $\pi$ in reverse order to have the most influential people early in the sequence.

The use of a priority queue, which takes $O(\log |P|)$ operations to pop the person with the smallest coverage confers Algorithm \ref{alg:min_cov} a computational cost of $O(|E| \log |P|)$.

\subsection{Maximum Coverage Iterative Peeling}
\label{sub:max_cov}
The Maximum Coverage Iterative Peeling Heuristic is a layered pruning strategy designed to iteratively sparsify $G$ by iteratively identifying and removing the individual with the highest coverage.
The heuristic iteratively identifies a \emph{layer} of people which cover all the tasks in $G$ and adds these people to $\pi$, in decreasing order of coverage.
Each such layer can be interpreted as a solution to the \emph{set cover problem}, as it consists of a subset of people whose collective edges cover all remaining tasks.

This heuristics is designed to achieve two simultaneous objectives: \emph{1)} sparifying $G$ by iteratively removing the person with the highest coverage, and \emph{2)} isolating low degree tasks, even those connected only to low degree people, early on in the process.

\subsubsection{Algorithm description}
We start with an empty list $\pi$ and perform iterative rounds of greedy removal.
In each round, we maintain a working copy of the graph and repeatedly select the person which covers the highest number of yet uncovered tasks.
This continues until all remaining tasks in the current round are covered.
Once coverage is complete, we remove the selected people and any now-isolated tasks, then begin a new round on the reduced graph.
The process continues until all the people have been processed.

\subsubsection{Implementation details}

\begin{algorithm}
\KwIn{a bipartite graph $G = (P,T,E)$}
\KwOut{Ordering $\pi$ of people}

$\pi \gets []$ \;
$N \gets |P|$ \;
\texttt{num\_inserted} $\gets 0$\;

\While{\texttt{num\_inserted} $<$ $N$}{
    $G.\text{remove\_nodes\_from}(\pi)$\;

    $\texttt{covered[$t$]} \gets 0\; \forall t \in T$ \;

    \texttt{num\_tasks} $\gets \left|\{G.\text{degree}(t) > 0,\; \forall t \in T \}\right|$\;
    \If{\texttt{num\_tasks} == 0}{\Return $[\pi | P]$}

    \texttt{num\_covered} $\gets 0$\;

    $\mathcal{Q}$ $\gets$ MinHeap()\;
    \ForEach{$p \in P$}{
        $\mathcal{Q}$.push($(-G.\text{degree}(p), p)$)\;
    }

    \While{\texttt{num\_covered} $<$ \texttt{num\_tasks}}{
        $(\texttt{priority}, p) \gets \mathcal{Q}.pop()$\;
        $\texttt{cov} \gets \left|\{{t \in G.\text{neighbors}(p)\; \textbf{if not}\; \texttt{covered[$t$]}}\}\right| $\;

        \If{$-\texttt{priority} \neq$ \texttt{cov}}{
            $\mathcal{Q}$.push($(-\texttt{cov}, q)$)\;
            \textbf{continue}\;
        }

        $\pi$.append($p$)\;
        \texttt{num\_inserted} $\gets \texttt{num\_inserted} + 1$\;
        \ForEach{$t \in G$\textnormal{.neighbors($p$)}}{
            \texttt{covered[$t$]} $\gets 1$\;
        }
        \texttt{num\_covered} $\gets$ \texttt{num\_covered} $+$ \texttt{cov}\;
    }
}
\Return{$\pi$}\;
\caption{Maximum Coverage}
\label{alg:max_cov}
\end{algorithm}

We adopt a priority queue with lazy updates, where people are prioritized by effective coverage multiplied by -1 in order to select the highest coverage at each round (Algorithm \ref{alg:max_cov}, lines 11---13). 
At the beginning, the coverage of each person corresponds with their degree.
Algorithm \ref{alg:max_cov} stops when all the people in $G$ have been added to $\pi$, which is initialized to an empty list.
At each round, the people inserted in $\pi$ at the previous round are removed from $G$ and a priority queue with the remaining people is initialized (lines 4---13).
Then, Algorithm \ref{alg:max_cov} pops from $\mathcal{Q}$ the person $p$ with the highest coverage and checks if the priority associated to $p$ is still updated (lines 15---17), in which case it adds $p$ to $\pi$ and marks the tasks to which $p$ is connected as covered (lines 20---24). 
If not, it updates the priority and pushes $p$ back in the queue (lines 17---19). This is a lazy update since people priorities are update only when needed.

Similarly to the \texttt{Minimum Coverage}, the computational complexity of Algorithm \ref{alg:max_cov} is $O(|E| \log |P|)$.

\subsection{Greedy $\tau(G)$ minimization: a boost for $\mathcal{P}(G)$}
\label{sub:reverse_greedy}
This heuristic is designed to find a removal order which approximate the bus-factor measure $\mathcal{P(G)}$.
It is conceptually similar to the \texttt{Minimum Coverage} heuristic, in that it works in reverse by adding to $\pi$ the person which contribute the least to the growth of $\tau(G)$.
In this way, the people who contribute the most to the growth of $\tau(G)$ are among the last to be added to $\pi$.
As with \texttt{Minimum Coverage}, the removal order is $\pi$ reversed.

The process can also be stopped when $\tau(G)$ reaches a certain threshold $t$ provided in input; in such a way the greedy $\tau(G)$ minimization can be combined with \texttt{Minimum Coverage} or \texttt{Maximum Coverage}, to obtain smaller values of $\mathcal{P}(G)$.

This heuristic is not appropriate to approximate $A(G, t)$, since growing $\tau(G)$ as slowly as possible might actually require to attach a person to a task in order to maintain $\tau(G)$ small.
Such a move, however, would cover all the tasks early one, maximizing the coverage.

\subsubsection{Algorithm description}
The heuristic receives the graph $G$ and a threshold $t$ in input and operates constructively.
First, it initializes a graph $\mathcal{U}$ that contains only the set of task nodes and no people. 
At each iteration, the person $p$ whose inclusion leads to the smallest possible increase in $\tau(G)$ is selected and added to $\pi$.
The person $p$ is then added to $\mathcal{U}$ with all their connections.
The process is repeated on the updated $\mathcal{U}$, until adding another person to it would make $\tau(G) > t$.

\begin{algorithm}
\KwIn{a bipartite graph $G=(P,T,E)$, a threshold $t \le |T|$}
\KwOut{Ordering $\pi$ of people}

$\mathcal{U} \gets \texttt{UnionFind(|T|)}$\;
$\Gamma[p] \gets [] \; \forall p \in P$\;
$\mathcal{Q} \gets \texttt{MinHeap()}$\;
$\pi \gets [] $\;

\ForEach{$p \in P$}{
        $(\text{roots}, \tau_p) \gets \mathcal{U}.\texttt{resulting\_size}(G.\texttt{neighbors($p$)})$\;
        $\Gamma[p] \gets \texttt{roots}$\;
        $\mathcal{Q}.\texttt{push}((\tau_p, p))$\;
}

\While{$\mathcal{Q}$ is not empty}{
    $(\texttt{priority}, p) \gets \mathcal{Q}.\texttt{pop()}$\;

    $\texttt{old\_roots} \gets \Gamma[p]$ \;
    $(\texttt{new\_roots}, \tau_p) \gets \mathcal{U}.\texttt{resulting\_size}(\texttt{old\_roots})$\;

    \uIf{$\tau_p > \texttt{priority}$}{
        $\mathcal{Q}.\texttt{push}((\tau_p, p))$\;
        $\Gamma[p] \gets \texttt{new\_roots}$\;
        \textbf{continue}\;
    }

    \If{$\texttt{priority} > t$}{
        \textbf{break}\;
    }

    $\mathcal{U}.\texttt{union}(\texttt{old\_roots})$\;
    $\pi.\texttt{append($p$)}$\;
}
\Return{$\pi$.\textnormal{reverse($\,$)}}\;
\caption{Greedy $\tau(G)$ Minimization}
\label{alg:min_tau}
\end{algorithm}

\subsubsection{Implementation details}
Greedily minimizing $\tau(G)$ at each step introduces significant computational challenges. 
In particular, evaluating the marginal increase in $\tau(G)$ caused by adding a person requires computing the connected components of $G$. 
A standard depth-first search (DFS) for connected components has a cost of $O(|P| + |T| + |E|)$. 
A naive implementation of the greedy heuristic would perform this DFS for each remaining person at each iteration.

As a result, the cost of one iteration is $O(|P|\cdot (|P|+|T|+|E|))$. Since this process is repeated over $|P|$ rounds (each adding one person), the overall time complexity becomes: $O(|P|^2 \cdot (|P|+|T|+|E|)) = O(|P|^3 + |P|^2|T| + |P|^2 |E|)$, which is prohibitive even for graphs of moderate size.

This is why in Algorithm~\ref{alg:min_tau} we use a union-find structure with path compression and union by rank to efficiently track the components of $G$ and $\tau(G)$.
Again, we use a lazily updated priority queue, where we prioritize people by their contribution to $\tau(G)$, and we stop the process where $\tau(G) > t$.
The function \texttt{resulting\_size($\cdot$)} simulates the addition of $p$ and their connections to $\mathcal{U}$ and returns the components to which $p$ would be connected after being added (\texttt{new\_roots}) and the resulting value of $\tau(G)$.

With the union-find structure, we can query $\tau(G)$ in $O(1)$ and perform a union operation in $O(\alpha(|T|))$, where $\alpha$ is the inverse Ackermann function which grows very slowly and can be treated as constant in practice.
Therefore, keeping track of the connected components through out the whole process requires $|E|$ union operations.
Therefore, the computational cost of Algorithm~\ref{alg:min_tau} is in the average case $O(|E| \log |P|)$.
However, if $G$ is dense the addition of a person might lead to the update of all the remaining people in $\mathcal{Q}$, requiring $O(|P|^2)$ updates.
As such, in the worst case, the computational complexity of Algorithm~\ref{alg:min_tau} is $O(|P|^2 \log |P|)$.

This is why we suggest to combine Algorithm~\ref{alg:min_tau} with one of the previously discussed heuristic, using a small threshold $t$ to approximate the bus-factor measure $\mathcal{P}(G)$.

\begin{figure}
    \centering
    \includesvg[width=\linewidth]{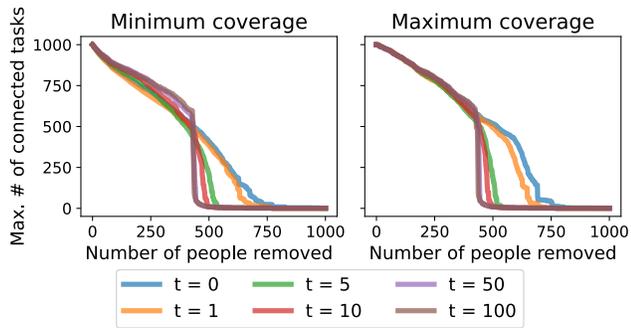}
    \caption{Threshold analysis for the greedy $\tau(G)$ minimization heuristic combined with \texttt{Minimum Coverage} and \texttt{Maximum Coverage}. $t=0$ means that the greedy $\tau(G)$ minimization heuristic has not been applied.}
    \Description{A figure which shows the effectiveness of combining the greedy tau minimization with the minimum coverage and maximum coverage heuristics.}
    \label{fig:threshold_analysis}
\end{figure}

We suggest to compute a partial order $\pi_r$ using Algorithm~\ref{alg:min_tau} with a small threshold (such as $t = 0.01 \times |T|$) and then using \texttt{Minimum Coverage} or \texttt{Maximum Coverage} on the graph $G = (\{P \setminus \pi_r\}, T, E)$ to compute the partial order $\pi_l$. The final order is then given by $\pi = [\pi_l | \pi_r]$.
In figure \ref{fig:threshold_analysis} we show the effectiveness of such a strategy.
We generated a bipartite power-law graph with Algorithm \ref{alg:bipartite_power_law_graph}, setting $|P| = |T| = 1000$, $\lambda_P = \lambda_T = 0.5$, and $k_P = k_T = 10$, and approximated the decay curve of $\tau(G)$ with both \texttt{Minimum Coverage} and \texttt{Maximum Coverage}, varying the threshold $t$.
Recalling that $\mathcal{P}(G)$ is the area under the decay curve of $\tau(G)$, we can observe that employing this mixed strategy pays off for both heuristics.

For $t = 0$, i.e. the original heuristic without the application of Algorithm~\ref{alg:min_tau}, we observe the largest area under the curve (0.4 for \texttt{Minimum Coverage} and 0.46 for \texttt{Maximum Coverage}).
As we increase the value of $t$, we can observe that the area under the curve becomes smaller.
For instance, for $t = 10$ we find an area under the curve of 0.35 for \texttt{Minimum Coverage} and 0.37 for \texttt{Maximum Coverage}.
However, increasing the threshold $t$ indefinitely does not necessarily yields improvements: for \texttt{Minimum Coverage} the area under the curve remains at 0.35 for both $t = 50$ and $t=100$, while for \texttt{Maximum Coverage} the area under the curve remains 0.36 for both $t = 50$ and $t=100$.
As such, selecting a too high threshold $t$ is likely to only require a higher computational time without yielding any benefit over smaller thresholds.

We finally note that for $t = |T|$ Algorithm~\ref{alg:min_tau} terminate when it has processed all the people in $P$; however, as it will be clear from our experiments in section \ref{sec:experiments}, this unbounded application of Algorithm~\ref{alg:min_tau} is inferior to the mixed strategy proposed here.

\section{Experimental evaluation}
\label{sec:experiments}

We conduct a comprehensive experimental study to evaluate the effectiveness, efficiency, and robustness of the proposed heuristics on large-scale bipartite graphs. 
While our techniques are motivated by theoretical insights, it is critical to assess their practical performance under realistic settings.
Our evaluation is guided by three main goals:
\begin{description}
    \item[Accuracy:] We assess the quality of each heuristic by measuring how closely they approximate the desired graph metrics, using a diverse benchmark of synthetic bipartite power-law graphs.
    \item[Scalability:] We measure runtime performance on graphs with millions of nodes and edges, verifying whether the heuristics scale efficiently and align with their theoretical complexity.
    \item[Sensitivity to Structure:] We analyze how heuristic behavior varies with key graph parameters, such as the skewness of degree distributions, to understand their adaptability across different input regimes.
\end{description}

We implemented our heuristics in pure Python 3.10 using the library \texttt{networkx} for graph handling and the standard module \texttt{heapq} for the priority queue implementation.
Furthermore, we implemented our own union-find structure with path compression and union by rank in Python, under just-in-time compilation through the library \texttt{numba}.
We ran the following experiments on a laptop with 32GB of RAM, and a 2.9GHz CPU.

\subsection{Accuracy Analysis}

\begin{table}
\centering
\caption{Performance per heuristic on the synthetic dataset, grouped by measure. Gap ratios (avg, min, max) are computed relative to the best strategy (lower is better).}
\label{tab:results_synthetic}
\begin{tabular}{llrccc}
\toprule
\textbf{Measure} & \textbf{Heuristic} & \textbf{\% 1st $\uparrow$} 
& \multicolumn{3}{c}{\textbf{Gap Ratio} $\left(\frac{s}{s^*}\right)$ $\downarrow$} \\
\cmidrule(lr){4-6}
& & & \textbf{avg} & \textbf{min} & \textbf{max} \\
\midrule
\multirow{5}{*}{\parbox{2.2cm} {\centering Avelino et al.\cite{avelino2016novel} $A(G,0.5)$}}
    & min\_cov   &  83.6 & 1.01 & 1.0 & 1.40 \\
    & max\_cov   &  6.6 & 1.05 & 1.0 & 1.21 \\
    & greedy$_I$   &  9.8 & 1.74 & 1.0 & 4.64 \\
    & degree   &  0 & 1.15 & 1.01 & 2.37 \\
    & \texttt{combined}   &  90.2 & 1.0 & 1.0 & 1.02 \\
\cmidrule(lr){1-6}
\multirow{5}{*}{\parbox{2.2cm} {\centering Piccolo et al.\cite{piccolo2024evaluating} $\mathcal{P}(G)$}}
    & min\_cov$_{\tau}$   &  78.7 & 1.0 & 1.0 & 1.03 \\
    & max\_cov$_{\tau}$   &  13.2 & 1.03 & 1.0 & 1.07 \\
    & greedy$_{\tau}$   &  8.1 & 1.05 & 1.0 & 1.15 \\
    & degree   &  0 & 1.09 & 1.04 & 1.28 \\
    & \texttt{combined}   &  91.9 & 1.0 & 1.0 & 1.01 \\
\bottomrule
\end{tabular}
\end{table}

Evaluating the quality of our heuristics poses some challenges: the underlying optimization problems are NP-hard, and optimal solutions are computationally infeasible to obtain on even moderately sized graphs. 
Moreover, no standard benchmark datasets exist for this specific task. 
To overcome this, we construct a large and diverse synthetic dataset that captures the skewed degree distributions typically observed in real-world projects \cite{piccolo2018design, yamashita2015pareto, majumder2019software, torchiano2011is}.

\subsubsection{Dataset generation}
We use Algorithm~\ref{alg:bipartite_power_law_graph} to generate 1,000 bipartite graphs using a power-law sampling process with parameters chosen uniformly at random from the following ranges:
\begin{align*}
    |P| &\in [1\,000, 2\,000] &|T| &\in [1\,000, 2\,000]\\
    \lambda_P &\in [0.3, 0.7] &\lambda_T &\in [0.3,0.7]\\
    k_P &\in [50, 300] &k_T & \in [50, 300]
\end{align*}

This randomized generation process avoids introducing systematic bias, while ensuring coverage of a wide spectrum of graph structures.

\subsubsection{Evaluation setup}
We evaluate heuristics on both $A(G, t)$, with $t=0.5$ as indicated by Avelino et al.\cite{avelino2016novel}, and $\mathcal{P}(G)$.
For $A(G, 0.5)$, we compare \texttt{Minimum Coverage}, \texttt{Maximum Coverage}, the removal by degree, and a baseline greedy heuristic that removes at each step the person isolating the most tasks (termed greedy$_I$).
For $\mathcal{P}(G)$, we compare \texttt{Minimum Coverage} and \texttt{Maximum Coverage} augmented with the greedy minimization of $\tau(G)$ (with $t=10$), the removal by degree, and the unbounded version of greedy $\tau(G)$ minimization (termed greedy$_\tau$).
We also consider a simple ensemble method: $\texttt{combined} = \min(\text{min\_cov}, \text{max\_cov})$.

Since these heuristics compute upper bounds, lower values are strictly better. 
To quantify relative performance, we compute: the percentage of graphs for which each heuristic returns the best estimate (i.e., lowest score), the average, minimum, and maximum gap ratios, where the gap ratio is defined as $s/s^*$ where $s$ is the heuristic's score and $s^*$ is the best score obtained across all heuristics for that instance.

\subsubsection{Results}
Table \ref{tab:results_synthetic} summarizes the results.
We observe that \texttt{Minimum Coverage} (min\_cov) is generally the best performing heuristic, finding the lowest value 83.6\% of times for $A(G,0.5)$ and 78.7\% of times for $\mathcal{P}(G)$. It also features the lowest average gap ratio.
\texttt{Maximum Coverage} (max\_cov) finds the best solution 6.6\% of times for $A(G,0.5)$ and 13.2\% of times for $\mathcal{P}(G)$, featuring the smallest maximum gap ratio for $A(G,0.5)$.
Notably, the degree heuristic never finds the best solution, and features the highest average gap ratio (with the exception of greedy$_I$ for $A(G,0.5)$).

The baseline greedy$_I$, despite finding the best solution 9.8\% of times, shows by far the worst gap ratio.
Also the heuristic greedy$_\tau$ does not impress on $\mathcal{P}(G)$, showing marginally better performance than the degree heuristic, despite its computational overhead.
The poor performance of these two greedy heuristics illustrates the difficulty of approximating the bus-factor as two natural greedy heuristics designed to address their specific objective function turn out to perform overall worse (or at best on par) than a generic heuristic such as the degree.

Finally, the \texttt{combined} pushes the performance even further and, besides the obvious increase in the percentage of times in which it finds the best solution, it also shows the complementarity between \texttt{Minimum Coverage} and \texttt{Maximum Coverage}: combined together, the two heuristics achieve a smaller maximum gap ratio than either of them alone.

\subsection{Running time analysis}

\begin{figure}
    \centering
    \includesvg[width=\linewidth]{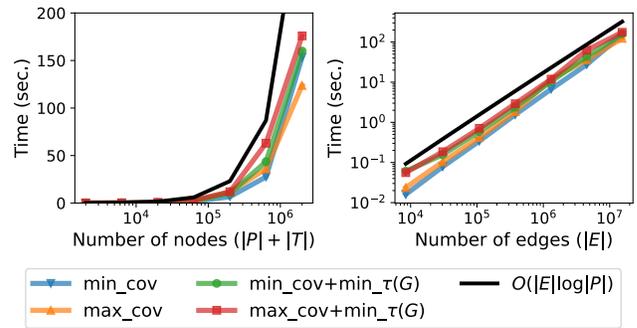}
    \Description{A plot of the running time of the heuristics developed in this paper, which show that our heuristics are fast and scalable.}
    \caption{Running time analysis. All the heuristics scale to very large graphs of millions of nodes and edges.}
    \label{fig:running_time_analysis}
\end{figure}

To evaluate the scalability of our heuristics, we conduct a series of timing experiments on a suite of synthetic bipartite graphs with increasing size. Specifically, we generate seven random bipartite graphs where the number of nodes $N$ for each node set (we set $|P| = |T| = N$) is logarithmically spaced in the range $[10^3, 10^6]$.
The connection probability is set to $p = \log(5N)/N$, ensuring that the expected degree grows slowly with the graph size and that the graph remains connected with high probability.

\begin{table*}
\centering
\caption{Correlation between graph structural properties and the gap ratio of each heuristic. A positive correlation indicates that the error (gap ratio) increases as the value of the structural feature increases. Conversely, a negative correlation coefficient implies that the gap ratio reduces as a the value of a feature increases. In bold, $|\rho| > 0.3$}
\label{tab:correlation_structure_error}
\begin{tabular}{lcccccccc}
\toprule
\multirow{2}{*}{\parbox{2.2cm}{\textbf{Structural \\ property}}} & \multicolumn{4}{c}{\textbf{$A(G,t)$}} & \multicolumn{4}{c}{\textbf{$\mathcal{P}(G)$}} \\
\cmidrule(lr){2-5} \cmidrule(lr){6-9}
 & min\_cov & max\_cov & greedy$_I$ & degree & min\_cov$_\tau$ & max\_cov$_\tau$ & greedy$_\tau$ & degree \\
\midrule
$|P|$  &   -0.01 & 0.14 & 0.23 &-0.02 &  0.11 & 0.03 & 0.06 & 0.04 \\
$|T|$  &   -0.03 & -0.09 & -0.22 & 0 &  -0.13 & -0.06 & -0.06 & -0.04 \\
$\lambda_P$  &   -0.01 & -0.03 & 0.06 & -0.09 &  -0.01 & 0.14 & 0.09 & 0.01 \\
$\lambda_T$  &   0.03 & -0.06 & -0.19 & -0.03 &  -0.04 & -0.17 & -0.03 & -0.14 \\
$k_P$ &   -0.27 & \textbf{0.41} & \textbf{0.49} & \textbf{-0.58} &  \textbf{-0.35} & \textbf{0.54} & \textbf{-0.31} & \textbf{-0.73} \\
$k_T$  &   0.21 & \textbf{-0.64} & \textbf{-0.62} & \textbf{0.34} & 0.22 & \textbf{-0.70} & \textbf{0.39} & \textbf{0.52} \\
$\delta(G)$  &  -0.09 & -0.15 & -0.27 & \textbf{-0.35} & -0.23 & -0.10 & 0.06 & -0.26 \\
$r(G)$  &  -0.17 & 0 & -0.17 & \textbf{-0.41} & \textbf{-0.52} & 0.02 & 0.03 & \textbf{-0.35} \\
\# leaf people  &  -0.22 & \textbf{0.65} & \textbf{0.85} & \textbf{-0.34} & -0.14 & \textbf{0.75} & \textbf{-0.40} & \textbf{-0.53} \\
\# leaf tasks &  0.29 & \textbf{-0.61} & \textbf{-0.56} & \textbf{0.68} & 0.29 & \textbf{-0.73} & \textbf{0.31} & \textbf{0.73} \\
\bottomrule
\end{tabular}
\end{table*}

We choose Erdős–Rényi-style random graphs for this experiment because, although they do not reflect the skewed degree distributions commonly found in real-world projects, they lack structural patterns that might be exploited by the heuristics, making them a robust baseline for measuring pure computational overhead.
The resulting graphs range from approximately $2,000$ to $2,000,000$ nodes and from about $8,400$ to over $15,000,000$ edges.
On each of these instances, we benchmark the execution time of \texttt{Minimum Coverage} and \texttt{Maximum Coverage} with and without the greedy $\tau(G)$ minimization, with threshold $t = 10$.

Results are shown in Figure \ref{fig:running_time_analysis}, reported as a function of the number of nodes and as a function of the number of edges.
We observe that the running time of all the heuristics is within $O(|E| \log |P|)$ (black line), in line with our theoretical analysis.
We also find that the base heuristics perform slightly faster than the heuristics combined with the greedy $\tau(G)$ minimization.
Finally, we note that on the largest graph, with $|P| = |T| = 1,000,000$ nodes and $|E| \approx 15,500,000$ edges, the heuristics have a running time in $[123, 176]$ seconds.

This shows that our heuristics are fast and scale to very large bipartite graphs. Considering that we wrote the code in pure Python without attempting to optimize its performance, we believe the running time can be improved further.

\subsection{Sensitivity to graph structure}

To understand how the structure of the input graph affects heuristic performance, we analyze the correlation between heuristic error and various structural features of the graph instances. 
Our analysis is conducted over the same $1\,000$ synthetic bipartite power-law graphs introduced earlier. Each graph is characterized not only by its generative parameters (e.g., degree distribution exponents and cutoffs), but also by derived structural features, including:
\begin{itemize}
    \item Graph density $\delta(G)$, defined as the ratio between the number of edges and $|P| \times |T|$;
    \item Number of leaf people and leaf tasks (nodes with degree 1);
    \item Degree assortativity $r(G)$, measured as the Pearson correlation between the degrees at both ends of each edge.
\end{itemize}
We compute Pearson correlation coefficients between these structural features and the gap ratio of each heuristic (defined as the ratio between the heuristic’s estimate and the best estimate among all heuristics). Table \ref{tab:correlation_structure_error} reports these correlations separately for both the $A(G, t)$ and $\mathcal{P}(G)$ measures.

\subsubsection{Structural robustness of \texttt{Minimum Coverage}}
The first notable result is that \texttt{Minimum Coverage} is largely insensitive to graph structure. 
For the $A(G,t)$ measure, its gap ratio never shows a correlation (in absolute value) greater than 0.3 with any structural feature, indicating strong robustness. 
For the $\mathcal{P}(G)$ measure, two moderate negative correlations appear with degree assortativity $r(G)$ (Pearson's $p=-0.52$) and with maximum person degree $k_P$ ($p = -0.35$).
This behavior aligns with prior findings in network robustness literature, where higher degree correlation is associated with increased robustness \cite{newman2003mixing, li2005towards, artime2024robustness, piccolo2025busfactor}.
The negative correlation with $k_P$ also reflects the fact that \texttt{Minimum Coverage} prioritizes high-degree nodes early in the removal sequence, leading to lower error in graphs with high-degree individuals.

\subsubsection{\texttt{Maximum Coverage} complements \texttt{Minimum Coverage}}
A comparison between \texttt{Minimum Coverage} and \texttt{Maximum Coverage} reveals complementary behavior. \texttt{Minimum Coverage} gap ratio is positively correlated with the number of leaf tasks and maximum task degree $k_T$, and negatively correlated with the number of leaf people and $k_P$.
\texttt{Maximum Coverage}, on the other hand, exhibits the opposite pattern: its error is positively correlated with $k_P$ and the number of leaf people, and negatively correlated with $k_T$ and the number of leaf tasks.

Furthermore, \texttt{Maximum Coverage} is uncorrelated with degree assortativity, which further differentiates it from \texttt{Minimum Coverage}. 
These findings suggest that while graphs rich in leaf tasks and skewed task degrees challenge \texttt{Minimum Coverage}, they are favorable to \texttt{Maximum Coverage}. 
Conversely, configurations with many leaf people hinder \texttt{Maximum Coverage} but not \texttt{Minimum Coverage}. 
This structural complementarity offers strong support for using the two heuristics jointly.

\subsubsection{Structure-Sensitive Behavior of greedy Heuristics}
The greedy heuristics show more sensitivity to structure: greedy$_\tau$ performs well when the number of leaf people and $k_P$ are high. 
In graphs where greedy$_\tau$ yields the best estimate, the number of leaf people is on average three times greater than that of leaf tasks, and $k_P$ is roughly twice $k_T$.
These conditions allow greedy$_\tau$ to effectively isolate components during node removal.

In contrast, greedy$_I$ struggles in such scenarios. 
The presence of many leaf people, each connected to a different task, limits the ability of the heuristic to isolate multiple tasks per iteration, leading to inflated bus-factor estimates. 
We find tht greedy$_I$ performs best in graphs with many leaf tasks and high task-degree skewness.
In its best-performing subset, the number of leaf tasks is on average seven times higher than the number of leaf people, and $k_T$ is more than twice $k_P$.

\section{Conclusions}
\label{sec:conclusions}
Understanding and mitigating project risks is a central concern in software engineering practice. 
In this paper, we addressed the problem of estimating the \emph{bus-factor} of software projects, a key indicator of a project's resilience to developer loss. 
We focused on two established formalizations from the literature, by Avelino et al.\cite{avelino2016novel} and Piccolo et al.\cite{piccolo2024evaluating, piccolo2025busfactor}, both of which pose computational challenges due to their NP-hardness.

To tackle this challenge, we modeled projects as bipartite graphs of people and tasks and introduced two novel peeling-based heuristics, Minimum Coverage and Maximum Coverage, to approximate the bus-factor. 
Our methods offer both theoretical scalability and practical effectiveness, outperforming the widely used degree-based heuristic, which we showed can significantly overestimate the bus-factor under realistic structural conditions.

We validated our approach through extensive experiments on synthetic datasets that model the skewed structure of real-world projects. 
Our heuristics consistently yielded more accurate bus-factor estimations and scaled to graphs with millions of nodes and edges within minutes.

This work opens the door to more robust and practical tools for project health analysis.
Looking ahead, we plan to explore even faster and more adaptive heuristics, and to integrate our algorithms into toolkits for mining software repositories and automated project monitoring tools. 
To encourage further research and practical adoption, we release our implementation as open-source software.

\begin{acks}
I am grateful to Mario Alviano, Pasquale De Meo, Marco Manna, Simona Perri, and Aldo Ricioppo for the fruitful conversations over the ideas underpinning this work.
Research partially supported by MUR under PNRR project grant number H23C22000860006, PE0000013-FAIR, Spoke 9 - Greenaware AI – WP9.2.
\end{acks}

\bibliographystyle{ACM-Reference-Format}
\bibliography{sample-base}

\end{document}